\documentclass{aa}  

\usepackage{graphicx}
\usepackage[varg]{txfonts}
\begin{document}

\title{Long-term photometric monitoring of the dwarf planet (136472) Makemake}
       
\author{
T. A. Hromakina\inst{1}
\and
I. N. Belskaya\inst{1}
\and
Yu. N. Krugly\inst{1}
\and
V. G. Shevchenko\inst{1}
\and
J. L. Ortiz\inst{2}
\and
P. Santos-Sanz\inst{2}
\and
R. Duffard\inst{2}
\and
N. Morales\inst{2}
\and
A. Thirouin\inst{3}
\and
R. Ya. Inasaridze\inst{4,5}
\and
V. R. Ayvazian\inst{4,5}
\and
V. T. Zhuzhunadze\inst{4,5}
\and
D. Perna\inst{6,7}
\and
V. V. Rumyantsev,\inst{8}
\and
I. V. Reva\inst{9}
\and
A. V. Serebryanskiy\inst{9}
\and
A. V. Sergeyev\inst{1,10}
\and
I. E. Molotov\inst{11}
\and
V. A. Voropaev\inst{11}
\and
S. F. Velichko\inst{1}  }

   \institute{Institute of Astronomy, V.N. Karazin Kharkiv National University, Sumska Str. 35, Kharkiv 61022, Ukraine\\
   \email{hromakina@astron.kharkov.ua}\\
    \and
        Instituto de Astrof{\'i}sica de Andaluc{\'i}a, CSIC, Apt 3004, 18080 Granada, Spain\\
        \and
        Lowell Observatory, 1400 West Mars Hill Road, Flagstaff, AZ 86001, USA\\
        \and
        Kharadze Abastumani Astrophysical Observatory, Ilia State University, K. Cholokoshvili Av. 3/5, Tbilisi 0162, Georgia\\
        \and
        Samtskhe-Javakheti State University, Rustaveli Street 113, Akhaltsikhe 0080, Georgia\\
        \and
        INAF -- Osservatorio Astronomico di Roma, Via Frascati 33, I-00078 Monte Porzio Catone (Roma), Italy\\
        \and
        LESIA -- Observatoire de Paris, PSL Research University, CNRS, Sorbonne Universit{\'e}s, UPMC Univ. Paris 06, Univ. Paris Diderot,\\
        Sorbonne Paris Cit{\'e}, 5 place Jules Janssen, F-92195 Meudon, France\\
        \and
        Crimean Astrophysical Observatory, RAS, 298409 Nauchny, Russia\\
        \and
        Fesenkov Astrophysical Institute, Observatory 23, Almaty 050020, Kazakhstan\\
     \and
     Institute of Radio Astronomy of the National Academy of Sciences of Ukraine, 4 Mystetstv St., Kharkiv, 61002, Ukraine\\
     \and
     Keldysh Institute of Applied Mathematics, RAS, Miusskaya Sq. 4, Moscow 125047, Russia\\  }

   \date{---;---}

 
  \abstract
   {}
   {We  studied the rotational properties of the dwarf planet Makemake.}
   {The photometric observations were carried out at different telescopes between 2006 and 2017.
    Most of the measurements were acquired in \textit{BVRI} broad-band filters of a standard Johnson-Cousins photometric system.}
   {We found that Makemake rotates more slowly than was previously reported.
        A possible lightcurve asymmetry suggests a double-peaked period of P = 22.8266$\pm$0.0001~h. 
        A small peak-to-peak lightcurve amplitude in \textit{R}-filter A = 0.032$\pm$0.005~mag implies an almost 
        spherical shape or near pole-on orientation.
        We also measured \textit{BVRI} colours and the \textit{R}-filter phase-angle slope and revised the absolute magnitudes.
        The absolute magnitude of Makemake has remained unchanged since its discovery in 2005.
        No direct evidence of a newly discovered satellite was found in our photometric data;
        however, we discuss the possible existence of another larger satellite.}
   {}

             
   \maketitle
%

\section{Introduction}

Dwarf planet (136472) Makemake is one of the largest (\textit{D}~$\sim$1400~km) and brightest
(geometric albedo \textit{p$_{\text{v}}$}~$\sim$0.8) known transneptunian 
objects (TNOs) \citep{Ortiz2012, Lim2010, Brown2013}.
Multiple spectral observations since its discovery in 2005 have revealed strong 
absorption bands of methane ice, which puts 
Makemake among only five methane ice-rich bodies in our solar system, together with
(134340) Pluto, (136199) Eris, Triton, and (90377) 
Sedna \citep[cf.][]{Licandro2006a, Tegler2008, Tegler2012, Brown2015, Lorenzi2015}.

The spectral slope of Makemake implies a somewhat reddish surface that could be 
explained by the presence of complex organic materials \citep{ Brown2007, Brown2015, Lorenzi2015, Perna2017}.
This makes Makemake's surface more similar to that of Pluto, rather than Eris with its more neutral spectral slope
\citep[e.g.][]{Licandro2006b, Alvarez-Candal2011, Merlin2015, Tegler2010, Tegler2012, Dumas2007}. 
However, unlike Pluto, according to rotationally resolved visible spectroscopy
it seems that the surface of Makemake is very homogeneous at the low spatial 
resolution achieved from the ground-based long-slit spectroscopy \citep{Perna2017}.

Polarimetric properties of Makemake are also similar to those of other large 
methane-dominated surfaces and differ from those of water-rich surfaces 
such as (136108) Haumea and (50000) Quaoar \citep{Belskaya2012}.

Several authors have performed photometric observations of Makemake in order to estimate its rotational period. 
The first attempt was made by \citet{Ortiz2007}, who suggested two possible 
values: 11.24~h and its double value of 22.48~h. 
Then, based on more precise observational data, a new value of 7.77~h 
was proposed by \citet{Heinze2009}. 
Finally, \citet{Thirouin2010} proposed a 7.7~h rotational period together
with its alias 11.5~h period, the former being more preferable.
The difficulties in determining Makemake's rotation period  are due to
a small lightcurve amplitude of 0.03 mag \citep{Heinze2009}. 
Robust characterisation of such small brightness variations requires very precise 
photometric measurements.  

Acquiring further photometric observations of Makemake is particularly important 
given the recent discovery of a Makemakean satellite \citep{Parker2016}. 
Although the influence on a rotation lightcurve from such a satellite is expected to be minimal, certain 
additional harmonics might be detected, which in turn could be used 
to constrain physical and orbital properties of the satellite. 
This discovery has also given a new interpretation on the thermal 
modelling results performed by \citet{Stansberry2008} and \citet{Lim2010}. 
The authors were able to fit Makemake's profile only while using a 
two-terrain model. The discovery of a moon may suggest that a
possible dark spot may correspond (at least partially)
to the satellite's surface and not to a certain dark area on Makemake.

We present a photometric study of the dwarf planet Makemake based on new observational data. 
A description of observations taken and data reduction is presented in Section~\ref{Obs}. 
In Section~\ref{Res} we show the results and analysis of photometric data,
which are followed by discussion and conclusions in Section~\ref{Dis}.

\section{Observations and data reduction}
\label{Obs}
        
The observations were carried out during 53 nights between 2006 and 2017.  
We used ten mid-sized telescopes at different observational sites, namely, the 3.6m Telescopio Nazionale Galileo (TNG),
the 2.6m Shain Telescope at Crimean Astrophysical Observatory (CrAO), 
the 2.5m Isaac Newton Telescope (INT) at Roque de los Muchachos Observatory, 
the 2.0m telescope at Peak Terskol Observatory (Terskol), 
the 1.5m telescope at Sierra Nevada Observatory (OSN), 
the 1.2m telescope at Calar Alto Observatory (CAO), 
the 1.0m Zeiss 1000 telescope at Simeiz Observatory (Simeiz), 
the 1.0m East and West telescopes at Tien Shan Astronomical Observatory (Tien Shan),
the 0.7m Maksutov meniscus telescope at Abastumani Astrophysical Observatory (AbAO), 
and the 0.7m telescope at Chuguev Observatory of V.~N.~Karazin Kharkiv National University (Chuguev).

The majority of data were acquired in 2012, 2015, and 2017. 
Table~\ref{table1} shows the information about the telescopes and the instruments,
as well as the total number of nights on each telescope and photometric filters that were used.
All the measurements were made using standard Johnson-Cousins photometric system 
in \textit{BVRI} broad-band filters or using no filters at all. 
Most of the observational data were obtained in \textit{R} filter.
Image reduction procedures were performed in a standard way which 
includes dark and/or bias subtraction and flat-field correction. 
The flat-field images were obtained during evening or morning twilight.

\begin{table*}
        \centering
        \caption{Summary of observational data}
        \label{table1}
        \begin{tabular}{p{1.6cm} p{0.7cm} c p{1.4cm} c p{1.5cm} c p{1.6cm} p{1.2cm} p{0.8cm}}
                \hline
                \hline
Obs. & D (m)&CCD camera & Number of pixels&Binning & Pixel scale ("/pxl) &Field of view& {Exp. time (sec)} &Filters&Nights\\
\hline
TNG$^{a}$&3.6&E2V 4240&2048~$\times$~2048&     1~$\times$~1            &0.252&8.6~$\times$~8.6      &90  &R&1\\
CrAO$^{b}$&2.6&FLI PL-4240&2048~$\times$~2048& 2~$\times$~2             &0.56&9.5~$\times$~9.5      &180  &R&3\\
INT$^{c}$&2.5&x4 EEV&2000~$\times$~4000&       1~$\times$~1            &0.33&11~$\times$~22         & 60&BVR&2\\
Terskol$^{d}$&2.0&FLI PL-4301&2048~$\times$~2048&1~$\times$~1              &0.31&10.7~$\times$~10.7   &180 &BVRI&2\\
OSN$^{e}$&1.5&CCDT150&2000~$\times$~2000&      2~$\times$~2              &0.46&7.8~$\times$~7.8     &400, 600&VR&16\\
CAO$^{f}$&1.2&DLR-III&4000~$\times$~4000&     1~$\times$~1               &0.314&21.5~$\times$~21.5  &300, 500&VR&8\\
Simeiz$^{g}$&1.0&FLI PL09000&3072~$\times$~3072&   3~$\times$~3          &0.56&9.5~$\times$~9.5     &240&R, Clear&8\\
Tian Shan$^{h}$&1.0&Apogee Alta F9000&3056~$\times$~3056& 1~$\times$~1    &0.74&18.9~$\times$~18.9  &300&R, Clear&4\\
AbAO$^{i}$&0.7&FLI IMG6303E&3072~$\times$~2048&  1~$\times$~1            &0.87&44.3~$\times$~29.5   &180 &Clear&6\\
Chuguev$^{j}$&0.7&ML47-10&1056~$\times$~1027&  1~$\times$~1              &0.95&16.8~$\times$~16.3   &240, 300 &R&3\\
                \hline
        \end{tabular}
        \begin{flushleft}
        $^{a}$Telescopio Nazionale Galileo, Roque de los Muchachos Observatory, Spain,
        $^{b}$Crimean Astrophysical Observatory, Ukraine,
        $^{c}$Isaac Newton Telescope, Roque de los Muchachos Observatory, Spain,
        $^{d}$Ukrainian-Russian Terskol Observatory, Russian Federation,
        $^{e}$Sierra Nevada Observatory, Spain, 
        $^{f}$Calar Alto Observatory, Spain,
        $^{g}$Simeiz Observatory, Crimea,
        $^{h}$Tien Shan Astronomical Observatory, Kazakhstan,
        $^{i}$Abastumani Observatory, Georgia,
        $^{j}$Chuguev Observatory, Ukraine.
        \end{flushleft}
\end{table*}

Aperture photometry of Makemake was performed using the \textsc{astphot} 
package developed at DLR (German Aerospace Center) by S.~Mottola \citep{Mottola1994}.
We used from three to five comparison stars in the object's field, which were inspected for possible variability.  
The radius of the photometry aperture was set using the full width at half maximum (FWHM) of
the seeing profile at each night.

The typical errors of the differential photometry were about 0.007-0.015~mag.
The accuracy of the Makemake's measured magnitudes in each filter are given in Table 2.

\section{Results and analysis}
\label{Res}
The observational circumstances and mean measured magnitudes of Makemake are shown in the Appendix (Table~\ref{table_data}).
The columns include mean UT, heliocentric (r) and geocentric ($\Delta$) distances, solar phase angle ($\alpha$), 
ecliptic longitude ($\lambda$) and latitude ($\beta$) in epoch J2000.0, mean reduced magnitude M(1, $\alpha$) 
and corresponding error, the filter in which the magnitude was measured, duration of observations ($\Delta$T), 
and finally, the telescope/observatory acronym. We note that for the nights when
only a few data points were acquired, the duration of observations is not shown in the table.

\begin{figure*}
        \includegraphics[width=\textwidth]{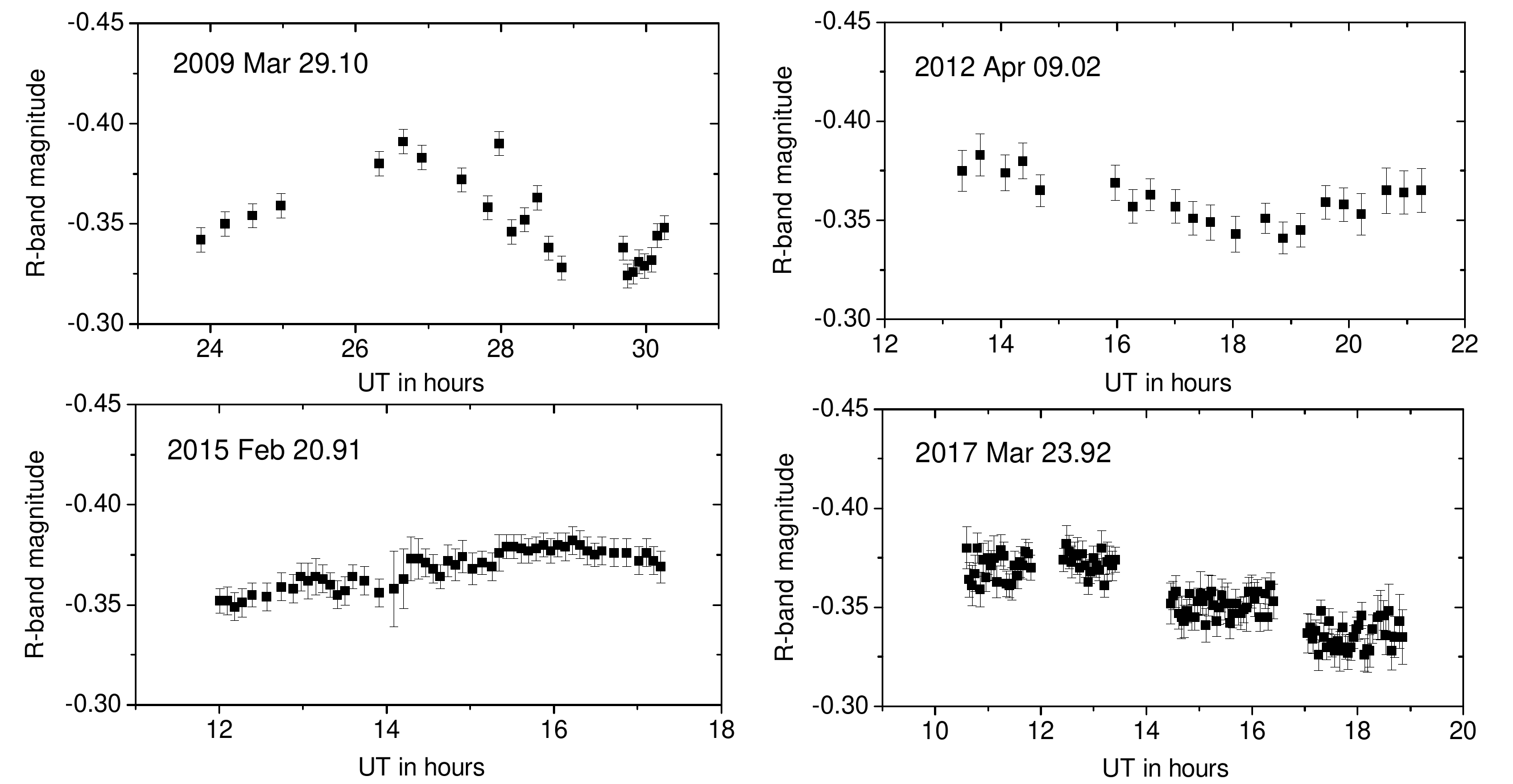}
        \caption{Examples of the individual nightly lightcurves acquired in R filter on different oppositions.}
        \label{best_LCs}
\end{figure*}

Examples of individual lightcurves from different oppositions are given
in Fig.~\ref{best_LCs}.
The amplitudes are small, but the lightcurve extrema can be clearly seen
within our accuracy of measurements.

\subsection{Search for rotational period}

We found that some of our long observations were inconsistent with a $\sim$7.7~h period, 
which is the preferred solution in the literature \citep{Heinze2009, Thirouin2010}.
In particular, this can be seen from the $\sim$8-hour  individual lightcurve, 
obtained on 23 March 2017 (Fig.~\ref{best_LCs}, lower right panel).

\begin{figure}
        \includegraphics[width=\columnwidth]{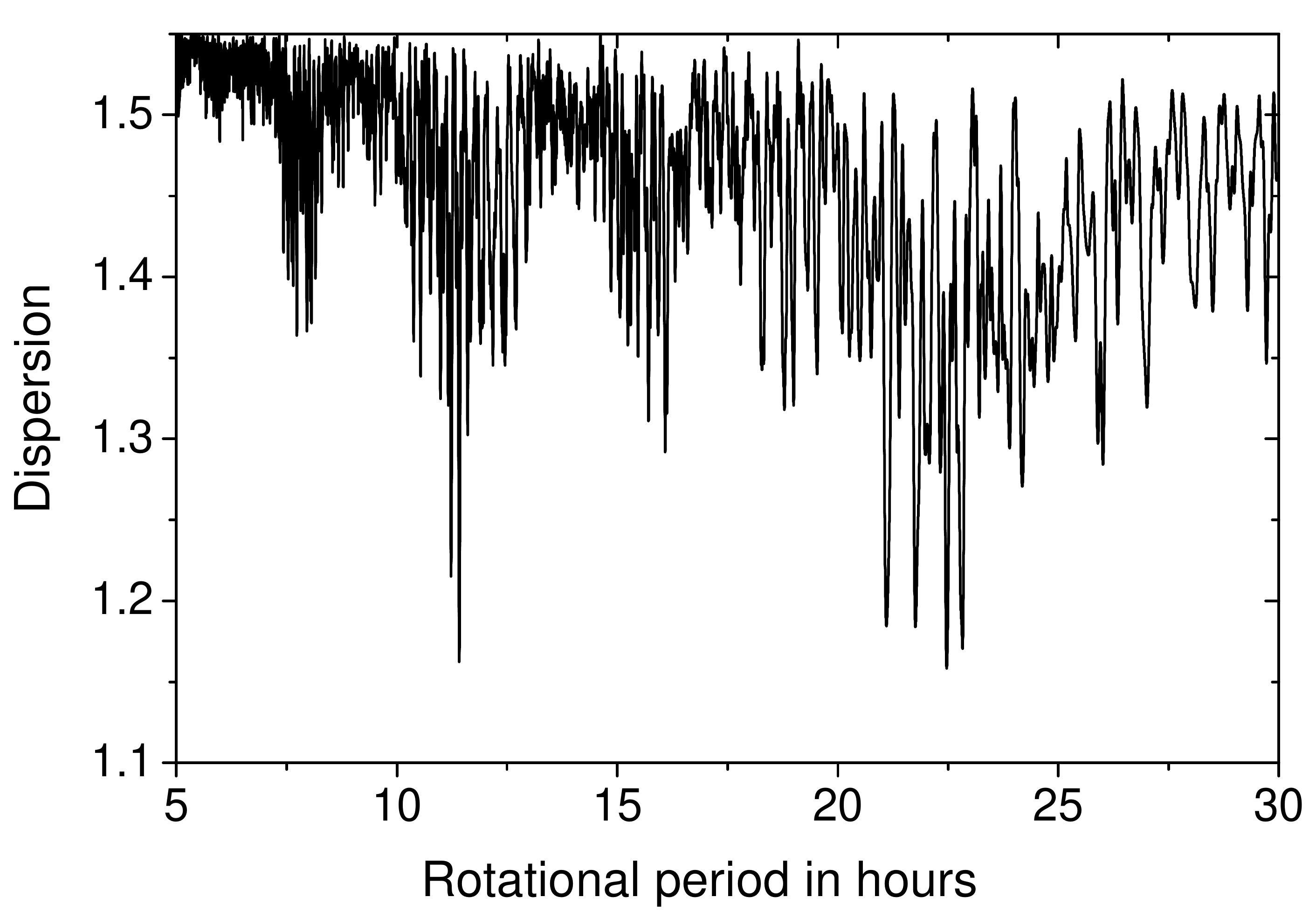}
        \caption{Resulting rotation spectrum that was acquired for the opposition in 2017.}
        \label{period2017}
\end{figure}

\begin{figure}
        \includegraphics[width=\columnwidth]{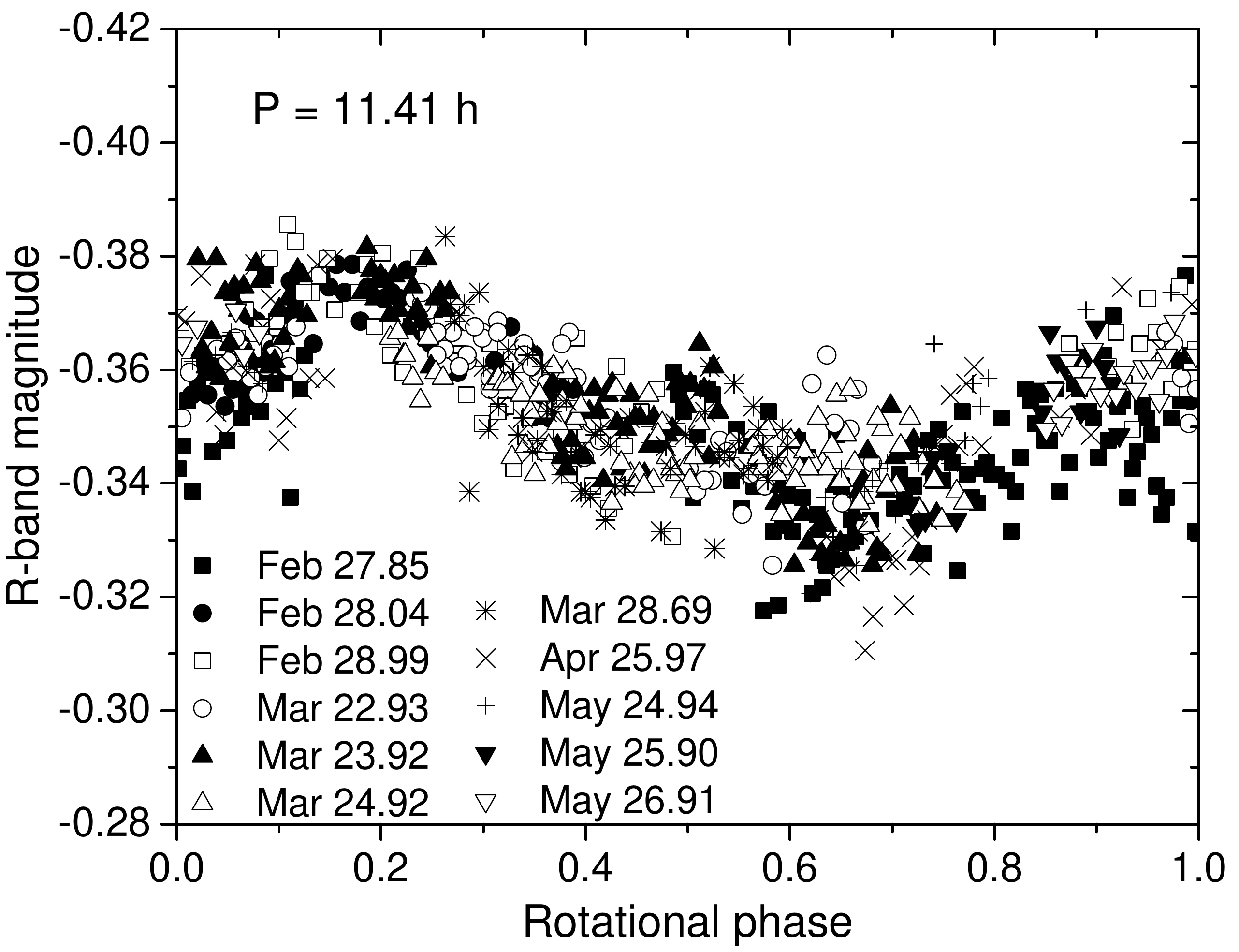}
        \caption{Composite lightcurve for the data from 2017 folded with a rotation period P=11.41~h. 
        Different symbols correspond to different dates.}
        \label{LC2017}
\end{figure}

We made a search for the rotational period following 
the method of Fourier analysis described by \citet{Harris1989}.
Specifically, a fourth-order Fourier function was used in the search. 
To derive the rotational period only \textit{R}-filter data were considered.
The probed periods were in the range from 5 to 30 hours and the step rate was equal to 1E-3 hours. 
For the initial search we used the data from a single opposition in 2017. 
The resulting rotation spectrum is presented in Fig.~\ref{period2017}.
It shows that the true period is around 11.41~h (or its double value), while the previously 
reported value around 7.73 h is connected with aliasing in the data.
The composite lightcurve for the 2017 data with a 11.41~h rotational period is presented in Fig. \ref{LC2017}.

\begin{figure}
        \includegraphics[width=\columnwidth]{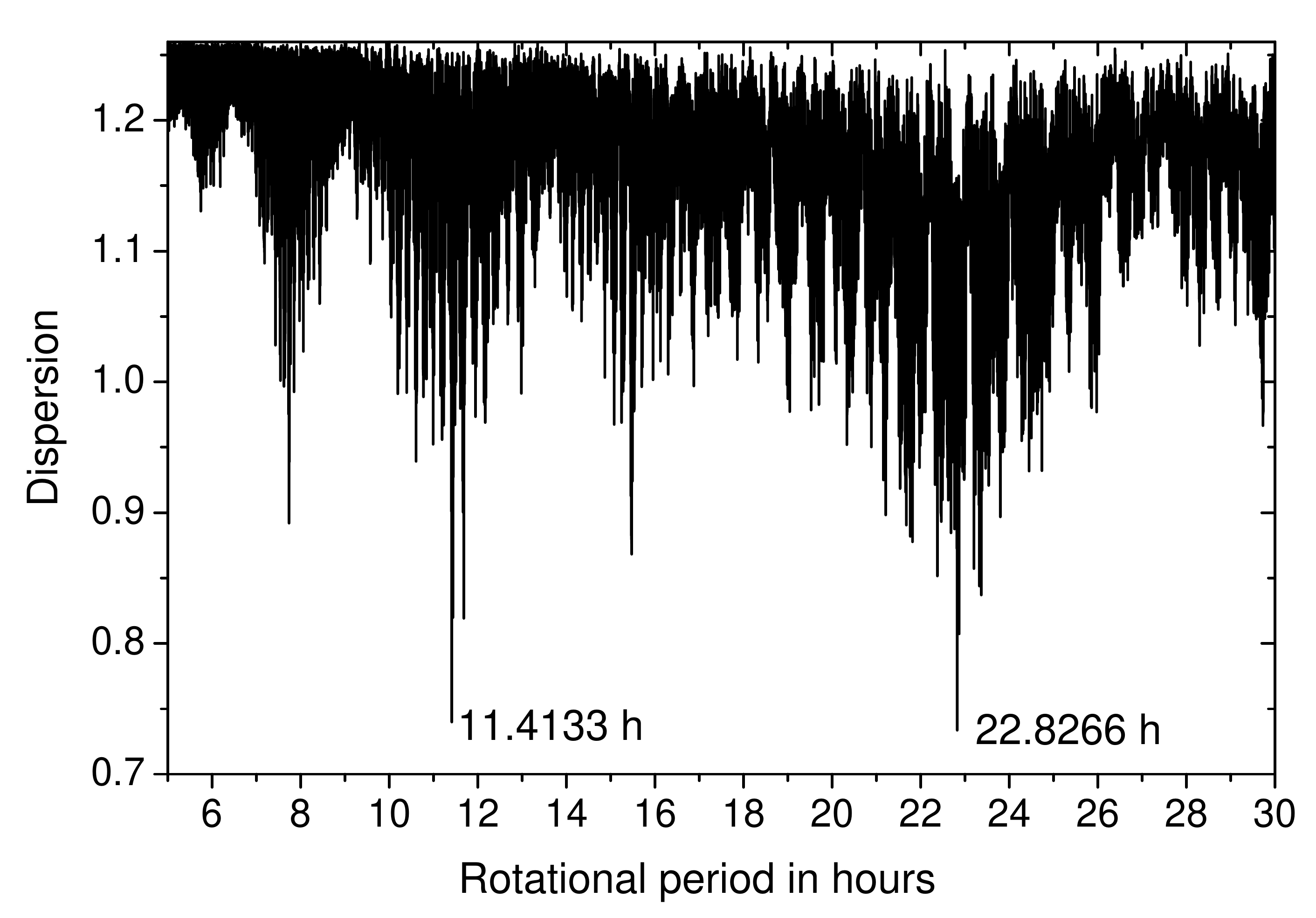}
        \caption{Resulting rotation spectrum  acquired with the data from different oppositions.
         The largest peaks correspond to 
         rotational periods P~=~11.4133~h (single-peaked) and P~=~22.8266~h (double-peaked).}
        \label{periodogram}
\end{figure}

\begin{figure}
        \includegraphics[width=\columnwidth]{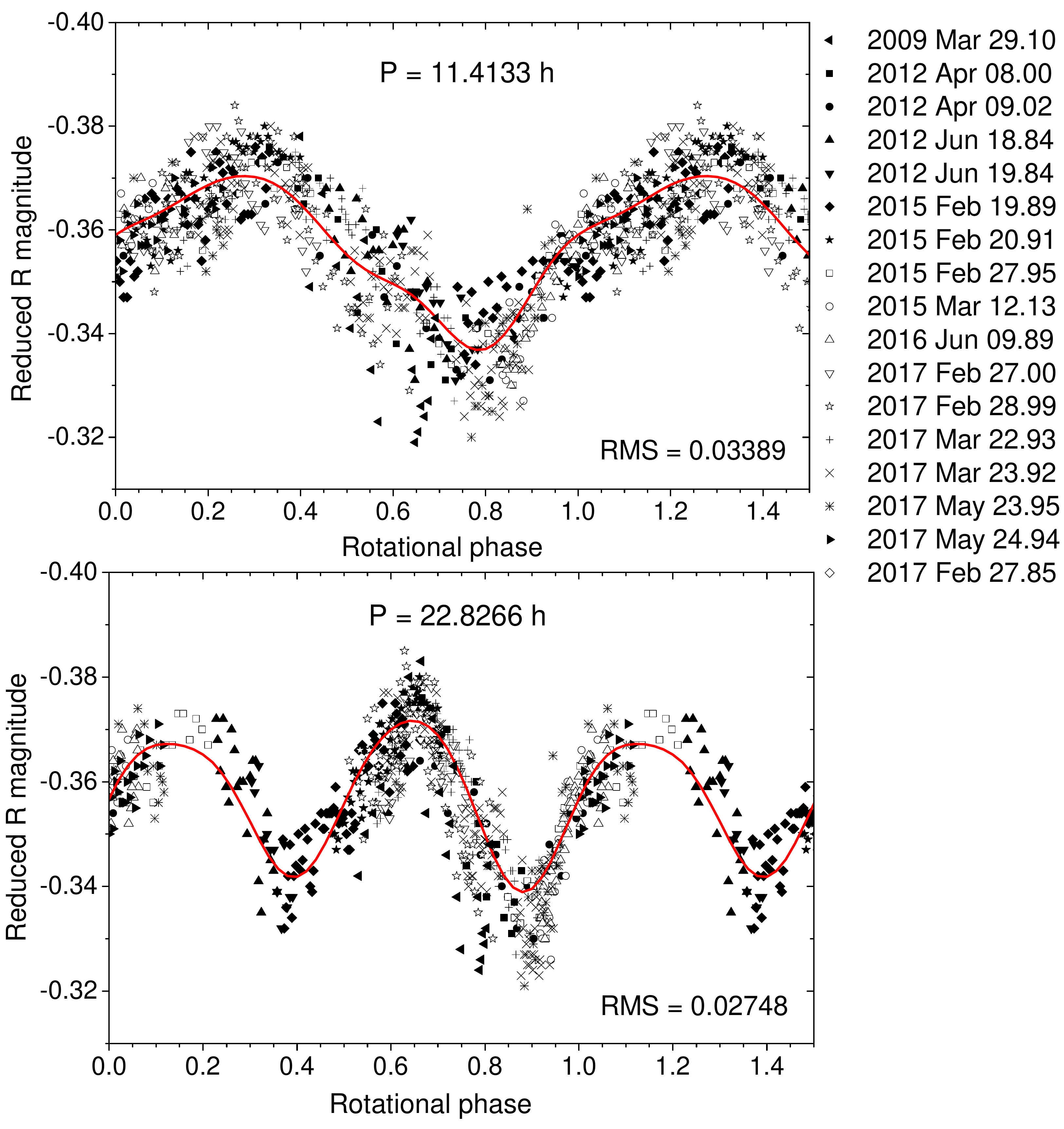}
        \caption{Composite lightcurves that are folded with rotational periods 
        P~=~11.4133~h (upper panel) and P~=~22.8266~h (lower panel). The solid line is a fourth-order Fourier fit.}
        \label{22h_our}
\end{figure}

Since Makemake's aspect of observations has changed very little over the past decade,
we combined data from several oppositions to determine a more precise value of the rotation period.
We used the data from 2009, 2012, and 2015-2017 when some long observations were acquired.
The new rotation spectrum with an increased step rate of 1E-5 hours is presented in Fig.~\ref{periodogram}.
Two definite dispersion minima at 11.4133~h and at 22.8266~h were found.
The amount of data from different oppositions was sufficient to have a good coverage for the long double-peaked period.
The composite lightcurves for both the single and double-peaked solutions
together with their Fourier fits are presented in Fig.~\ref{22h_our}.
The lightcurve with the period P~=~22.8266~h has a slightly lower RMS than that with the period P~=~11.4133~h.
Naturally, the fit is  better only because there is less data overlap.
Alternatively, this difference can also be due to the possible 
lightcurve asymmetry which is analysed in Sect. 3.2.

\subsection{Analysis of the lightcurve behaviour}

In the case of Makemake with its small amplitude, both single and double-peaked periods are possible (see section~\ref{Dis} for the discussion), but the existence of a lightcurve asymmetry would be  good evidence of a double-peaked lightcurve.
For the analysis we used only those observations that  covered a time span of more than 4 hours 
and had photometric errors <0.01 mag (cf. Table A.1.)

\begin{figure}
        \includegraphics[width=\columnwidth]{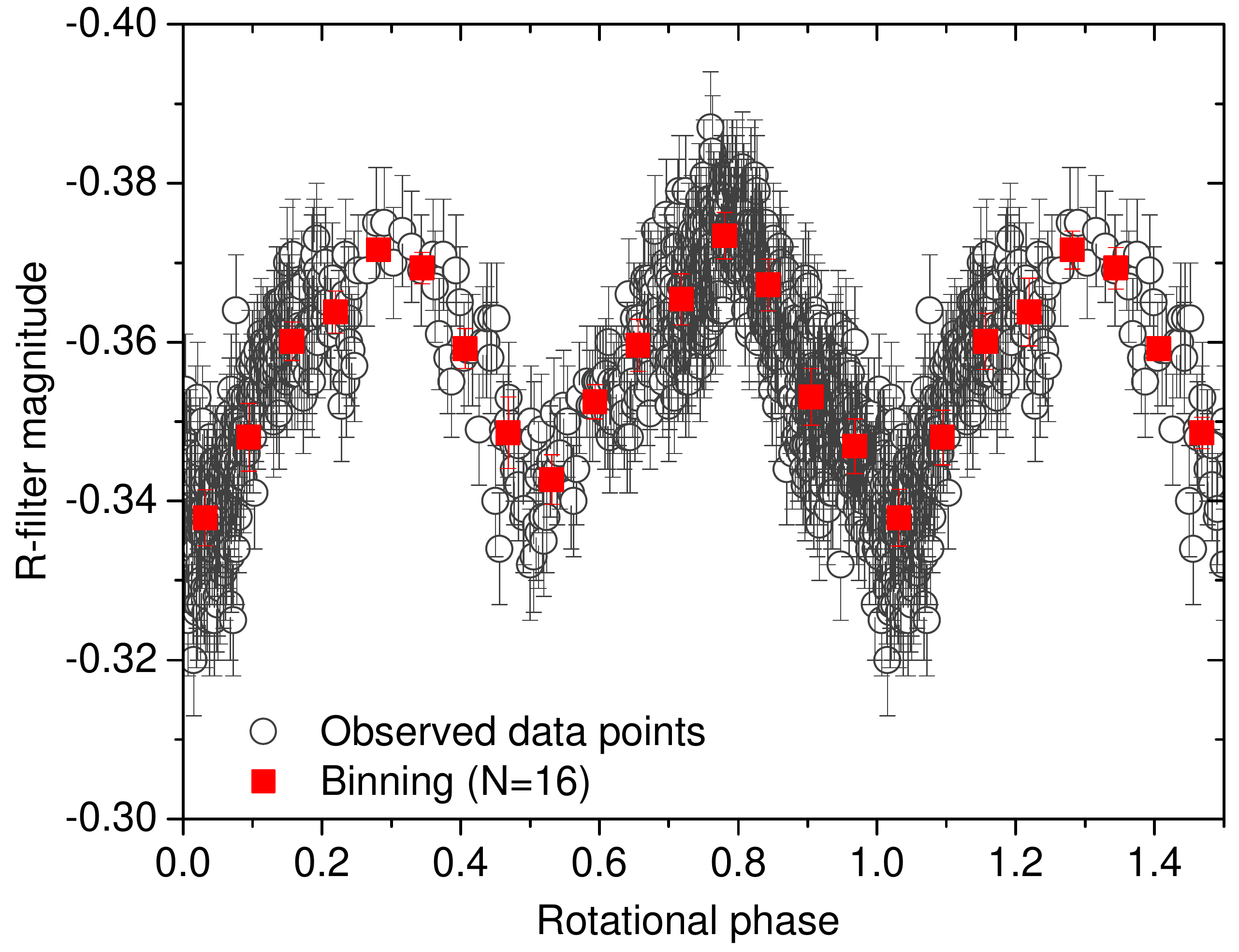}
        \caption{Composite lightcurve using a period of 22.8266~h and binning with N=16.}
        \label{Binning_figure}
\end{figure}

The lightcurve for a 22.8266~h period showed certain signs of asymmetry, 
as can be seen in Fig.~\ref{Binning_figure}: one maximum looks sharper, or more angular, than the other.
In order to estimate the level of significance of such an asymmetry, we performed a binning analysis.
The lightcurve was binned by calculating the average of data points that fall into each binning area.
The binning was done using an even number of bins, so for each $j$-th bin $b_j$ in the first half of a lightcurve there would be a corresponding bin $b_{j+N/2}$ in the second half, where $N$ is a total bin count.
To obtain the significance of a difference between two parts of a lightcurve,
we calculated the Student’s t-test value as
\begin{equation}
t = \sum_{j=1}^{N/2} \frac{\mid{b_{j+N/2} - b_j}\mid} {s \sqrt{B_j^{-1} + B_{j+N/2}^{-1}}},
\end{equation} 
where
\begin{equation}
s = \sqrt{\frac{ (B_j - 1) \delta b_j^2  + (B_{j+N/2}) \delta b_{j+N/2}^2 }  {B_j + B_{j+N/2} - 2}  };
\end{equation} 
 $\delta b_j$ and $\delta b_{j+N/2}$ are the uncertainties of the corresponding bin values, 
which were calculated as a standard deviation from the average of the real data points that 
fall into the $j$-th and $i+N/2$-th binning sections, respectively;  and $B$ is the total number of points in the bin.

For $N$ in the range from 10 to 30 the existence of an asymmetry is confirmed at a confidence level of 95{\%}.
Thus, we consider the long double-peaked rotational period to be more likely, 
although the single-peaked solutions cannot be completely discarded.

\subsection{Coincidence of the long rotation period with literature data}

In order to further investigate our finding of a long double-peaked 
rotational lightcurve we made use of literature values published by \citet{Heinze2009},
which were the only data obtained with a very good precision (around 0.01 mag) 
The authors indicated the 11.4~h period as an alias.

\begin{figure}
        \includegraphics[width=\columnwidth]{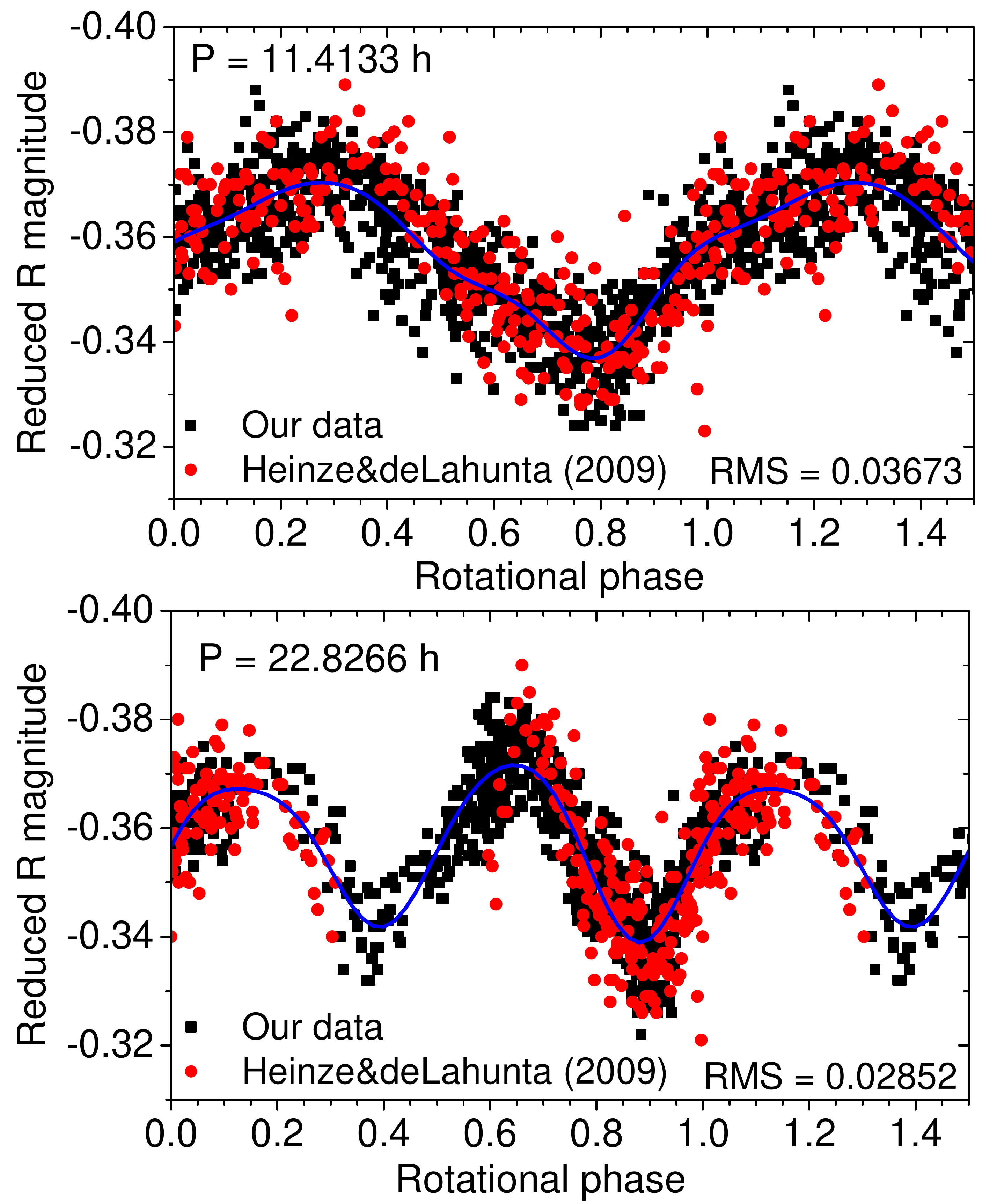}
        \caption{Composite lightcurves obtained using our data and literature values with
        the rotational periods P~=~11.4133~h (upper panel) and P~=~22.8266~h (lower panel).
        The solid line is a fourth-order Fourier fit.}
        \label{11h_heinze09}
\end{figure}

The composite lichtcurves with the rotational periods of 11.4133~h and 22.8266~h 
that were made using our and literature data are presented in Fig.~\ref{11h_heinze09}. 
Both periods are consistent with literature data.
In particular, in the data from \citet{Heinze2009} we notice a similar type of  lightcurve behaviour
for the double-peaked lightcurve as seen in our data. 

\begin{figure}
        \includegraphics[width=\columnwidth]{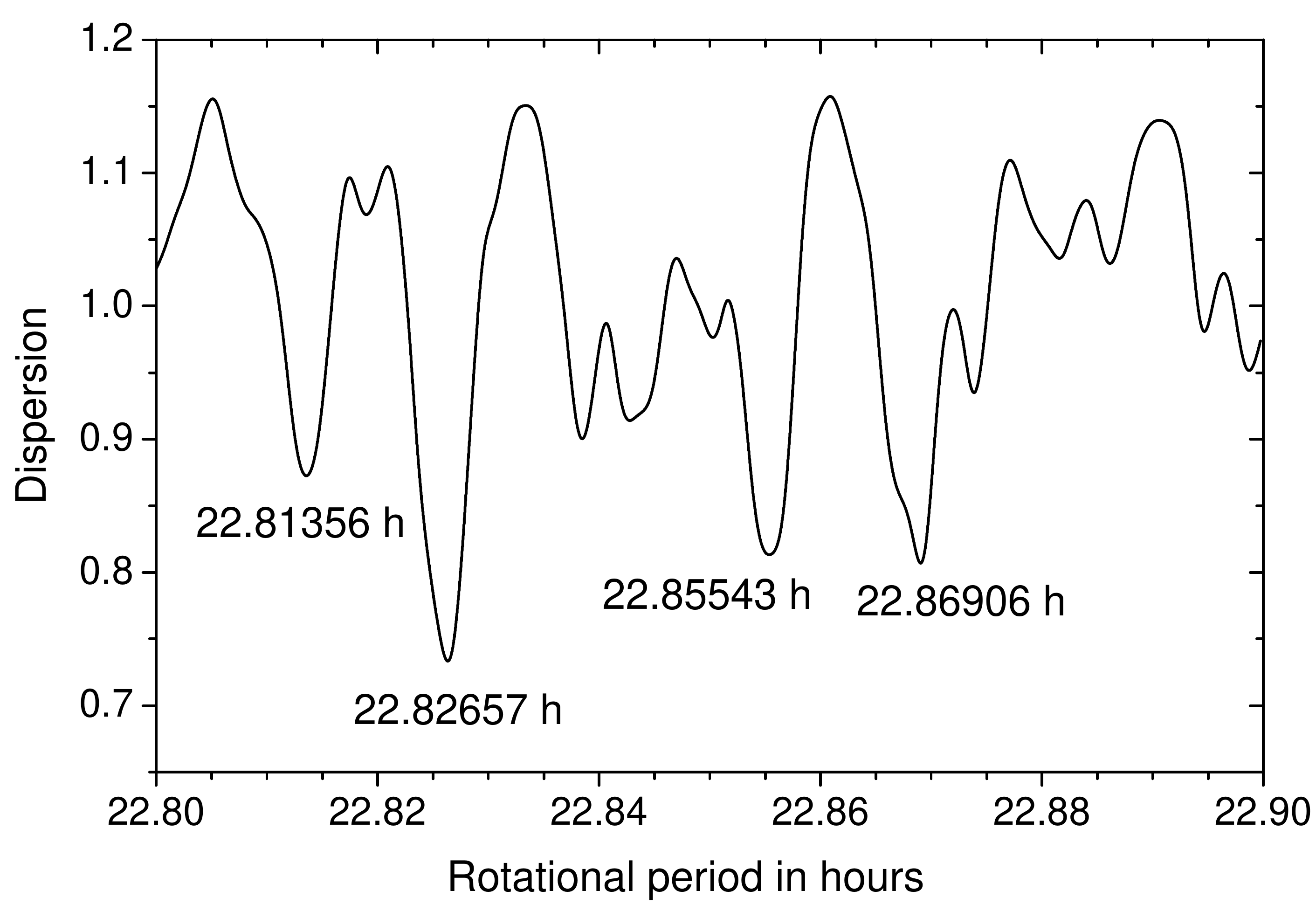}
        \caption{Rotation spectrum that was acquired in the area around $\sim$22.82~h using our data and literature values.}
        \label{Prdgrm}
\end{figure}

In order to increase the precision of a rotational period, we performed a period search 
using our data and the literature values in the area around $\sim$22.82~h with a step size of 1E-5 hours. 
The rotation spectrum is shown in Fig.~\ref{Prdgrm}. 
The minimal dispersion `noise spectrum' corresponds to the same period value 
that was already found during the search using only our data.

As was already mentioned, the data that was used for the rotation period determination was acquired in \textit{R} filter. 
However, on April 8-9, 2012, long lightcurves in both \textit{V} and \textit{R} filters were obtained. 
We have not found any significant differences between them.
Also, we did not find any difference in our \textit{R}-band data and \textit{V}-band data 
from \citet{Heinze2009}. In all filters Makemake has extremely low peak-to-peak lightcurve 
amplitude of $\sim$0.03~mag and shows the same lightcurve features.

Hence, we conclude the value of rotational period is P~=~22.8266$\pm$0.0001~h.
The calculated peak-to-peak lightcurve amplitude using a Fourier fit is 
A~=~0.032$\pm$0.005~mag.
The high accuracy of the rotational period is achieved by the long time span of observations 
(around 10 years including the literature data). The uncertainty was determined by 
 changing the found period until all of the data in the composite lightcurve still fits, and by  using a more formal estimation that depends on the total number of rotational cycles ($N$) during a given period of time. 
When using the first method the noticeable mismatches in the composite lightcurve 
started appearing with a shift of less than 0.0001 h.
In the second case, the accuracy can be found as the relation of $\Delta t/N$, 
where $\Delta t$ is the uncertainty of the time distance between two consequent extrema, 
which depends on the accuracy of measurements and on the sharpness of the extrema. 
For our data we can safely assume that this value is within half an hour.
Then, for P = 22.8266 h the error will also be  around 0.0001~h.

Makemake is large enough to be in hydrostatic equilibrium and to have an oblate Maclaurin spheroid shape (a=b>c). 
With the discovery of a satellite on the edge-on orbit the near equator-on aspect of observation became more feasible than the pole-on orientation \citep{Parker2016}.
In this case, Makemake's lightcurve variations are more likely caused by surface heterogeneity.

\subsection{Magnitude phase dependence and colour indices}

From our multi-colour observations we were able to determine the mean surface colour indices of Makemake.
In order to account for possible surface albedo variations,  colours were first
calculated using  almost simultaneously acquired data during only  one night. 
We were not able to detect any colour variations within 
the uncertainties, thus we report here the averaged values when multiple observations were available.
The measured surface colours for Makemake are
\textit{B-V} = 0.91$\pm$0.03~mag, 
\textit{V-R} = 0.41$\pm$0.02~mag, and
\textit{V-I} = 0.65$\pm$0.03~mag.
Our results are in agreement with previously reported values \citep{Rabinowitz2007, Jewitt2007}.
Thus, we can confirm a reddish surface similar to that of Pluto. 
This is consistent with the findings of \citet{Perna2010} that both Pluto and Makemake belong to the
BR taxonomic class, whereas the surface of Eris is more neutral and was classified in the BB taxon.

\begin{figure}
        \includegraphics[width=\columnwidth]{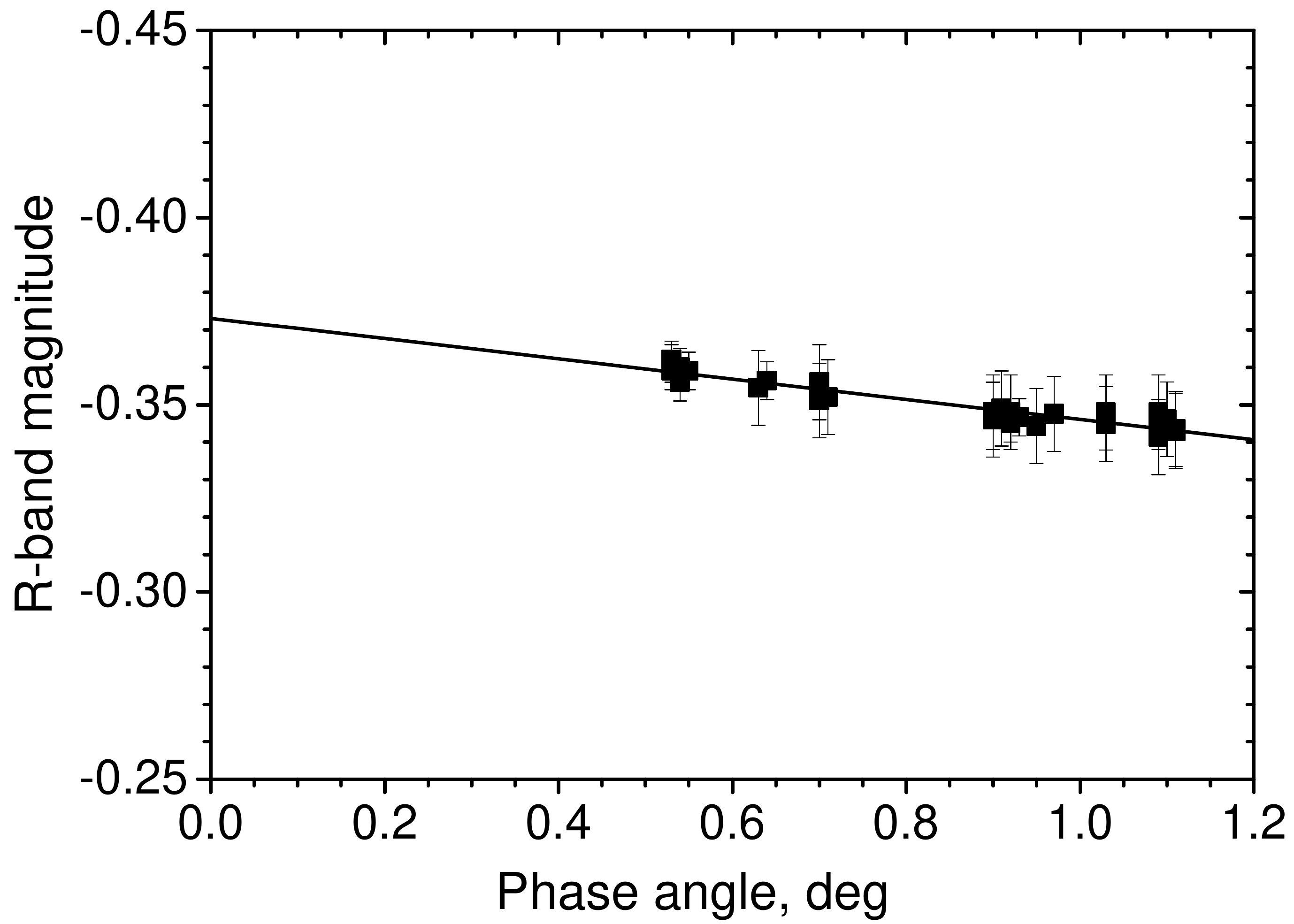}
        \caption{Magnitude phase dependence in R filter.}
        \label{slope}
\end{figure}

The phase-angle dependence of Makemake's magnitude was measured in the phase angle range of 0.5-1.1$^{\circ}$. 
It is the largest angle range that available for observations from Earth since the discovery of Makemake in 2005.
We have taken into account the lightcurve variations even if they are small.
The magnitude phase dependence of Makemake is presented in Fig.~\ref{slope}.
We found a linear slope of the phase dependence to be 0.027$\pm$0.011~mag/deg in \textit{R} filter. 
This value of a phase slope is slightly smaller than previously determined values for the same phase angle range
in \textit{V} filter: 0.037$\pm$0.013 mag/deg \citep{Heinze2009}, 
and 0.054$\pm$0.019 mag/deg \citep{Rabinowitz2007}.
No opposition surge was seen in our data. 
We assume that the opposition surge of high-albedo Makemake is very narrow, starting 
at  phase angles of less than 0.5$^{\circ}$, which are not covered by our observations.
Furthermore, \citet{Belskaya2003} showed that very narrow opposition surges seem to be typical for TNOs.
For example, an opposition effect (at phase angles of less than 0.1$^{\circ}$) was found at Triton, Neptune's satellite \citep{Buratti2011}.

\section{Discussion and conclusions}
\label{Dis}

Photometric variability of small  solar system bodies is most often caused by aspherical shape, surface albedo variations, or  binarity.
Single-peaked lightcurves are typically associated with albedo variations, and double-peaked lightcurves with elongated shape.
Quite frequently it is hard to distinguish between these two cases, and additional information is needed. 

In the case of distant solar system objects, such as Makemake, this is a particularly challenging task.
Firstly, the body could be near its polar aspect and hence have very small lightcurve amplitude. 
In the near-equatorial aspect a small lightcurve amplitude undoubtedly implies an almost spherical shape for the body.
An example of an object with a polar aspect is the New Horizons flyby target 2014 MU69. 
From a small lightcurve amplitude, \citet{Benecchi2018} suggested that the object is either nearly spherical 
or its polar axis is oriented towards the line of sight to Earth. 
The recent close up observations confirmed the pole-on orientation 
and revealed that 2014 MU69 is a bi-lobate contact binary \citep{Stern2019}.

From the analysis of our data and the literature values we discovered a possible asymmetry in the photometric lightcurve.
The existence of asymmetry suggests that the most probable cause of brightness variability would be 
shape irregularities and/or surface variations of albedo. 
In the case of the dwarf planet Ceres, an asymmetrical double-peaked lightcurve with a small amplitude of $\sim$0.03~mag
is primarily caused by albedo variations \citep{Chamberlain07, Reddy15}.

A previous study of Makemake's spectral data suggests that its surface is quite rotationally
homogeneous \citep{Perna2017}, although these data cannot pinpoint variations of $\sim$3{\%}. 
Moreover, neither photometric nor spectroscopic observations can
detect latitude variations, if such are present on Makemake's surface.

Lightcurve asymmetry due to shape can be explained by surface topographic features.
However, taking into account  Makemake's size and assuming a range of possible densities by varying ice/rock ratio, 
\citet{Rambaux2017} argue that a possible mountain on Makemake cannot be higher than 10~km. 
Such a relatively small feature (assuming that its albedo is not different 
from the rest of the surface) would give very little input of less than 0.001~mag into the brightness lightcurve. 
The observed amplitude difference for Makemake, on the other hand, reaches $\sim$0.01~mag.
From this we can assume that both variability causes might be present on Makemake: 
small albedo variations that were not detected from spectral 
observations together with minor deviations from the symmetrical shape.

The first observations of the Makemakean satellite by \citet{Parker2016} were quite sparse and it was therefore not possible
to determine its orbit and consequently the total mass of the Makemake plus satellite system.
Using the known magnitude difference between the primary and secondary, and assuming the lowest possible 
albedo for solar system objects of 4{\%}, the upper limit of the satellite's diameter would be $\sim$100~km. 
A satellite of this size can decrease the total brightness by about 0.01~mag.
From our data we could not find any effect on a lightcurve from the satellite. 
Given such a small input and an orbital period of more than $\sim$12~days 
the chance of a confident detection of a satellite influencing the rotational lightcurve from a mid-sized telescope is rather small. 

The slow rotational period of Makemake can be caused by the tidal effects between the primary and secondary body.
It was shown by \citet{Thirouin2014} that binary bodies tend to have longer rotational periods.
The discovered satellite lacks sufficient mass to have slowed down Makemake to the current slow rotational period.
Moreover, \citet{Parker2016} argues that the known Makemake satellite can
partially account for the dark area needed to fit the thermal observations by \citet{Lim2010}.
However, it can account for only about 1{\%} of the dark terrain 
and the rest of the area should correspond to Makemake's surface or to another, as yet undiscovered, larger dark satellite.

In this regard, we can consider the possibility that Makemake's photomeric variability is due to the existence of one more satellite. 
Using the formalism from \citet{Descamps2008} in order to slow down Makemake's rotational period to 22.8~h, the satellite
should reside at a distance of  5000~km  and have a mass ratio of 0.03 with respect to the primary body. 
This would give a specific angular momentum of 0.14.
Depending on its density, the size of such a satellite would be of the order of 400~km in diameter
and its area would be more than 9{\%} of Makemake's total area. 
At this distance from the primary body, the satellite would be outside the Roche limit.
Assuming 1500~kg~m$^{-3}$ density of Makemake, tidal locking would occur at about 3000~km distance
from the primary, still away from the Roche limit, but the mass ratio of the satellite to Makemake would
have to be higher than 0.05 to have slowed down Makemake's rotation from a primordial spin to 22.8~h. 
In this case the required size of the putative satellite should be at least 550~km. 
By area, such an object would have more than 15{\%} of Makemake's surface.
Hence, such an undiscovered satellite could also potentially explain the need for two-albedo terrains in the thermal modelling.
Notably, it will be close enough to Makemake so that it would not be detectable even with the
current space telescopes or the large ground-based ones \citep[e.g.][]{Brown2006}.

If such an undiscovered satellite exists and has an irregular shape, it could induce periodic variations 
of a small amplitude in Makemake's rotational lightcurve. 
The satellite would have to be outside the hydrostatic equilibrium, and consequently could be responsible for the detected
photometric lightcurve. Hence, the existence of an undiscovered satellite would have slowed down Makemake's rotation,
provide enough dark terrain to explain the two-terrain model needed by the thermal data.
It could also explain the double-peaked nature of the lightcurve without requiring
an asymmetry in Makemake's shape.
Such a close-in satellite might be discovered during a stellar occultation,
but the non-detection of any satellite during the 2011 occultation observed 
by \citet{Ortiz2012} does not rule it out, as the object could easily 
have been located north of the S. Pedro de Atacama chord.

Our sidereal rotation period measurement  was determined with enough precision to allow us to find
the rotational phase during the occultation event observed by \citet{Ortiz2012}. 
We found that the occultation event happened when Makemake was near  its maximum brightness.
Using the phase-angle slopes that are reported in this paper, the brightness at the moment 
of occultation, and assuming the absence of an opposition effect, we found new values of  
the absolute magnitudes to be \textit{H$_{\text{V}}$}=0.049$\pm$0.02~mag and 
\textit{H$_{\text{R}}$}=-0.388$\pm$0.02~mag in \textit{V} and \textit{R} filters, respectively.
We used those values together with an equivalent diameter of Makemake found from occultation 
to recalculate Makemake's geometric albedo. The revised albedo values are
\textit{p$_{\text{v}}$}=0.82$\pm$0.02 in \textit{V} filter and \textit{p$_{\text{r}}$}=0.89$\pm$0.02 in \textit{R} filter.
This result is more similar to that proposed by \citet{Brown2013} and \citet{Lim2010}.

\begin{figure}
        \includegraphics[width=\columnwidth]{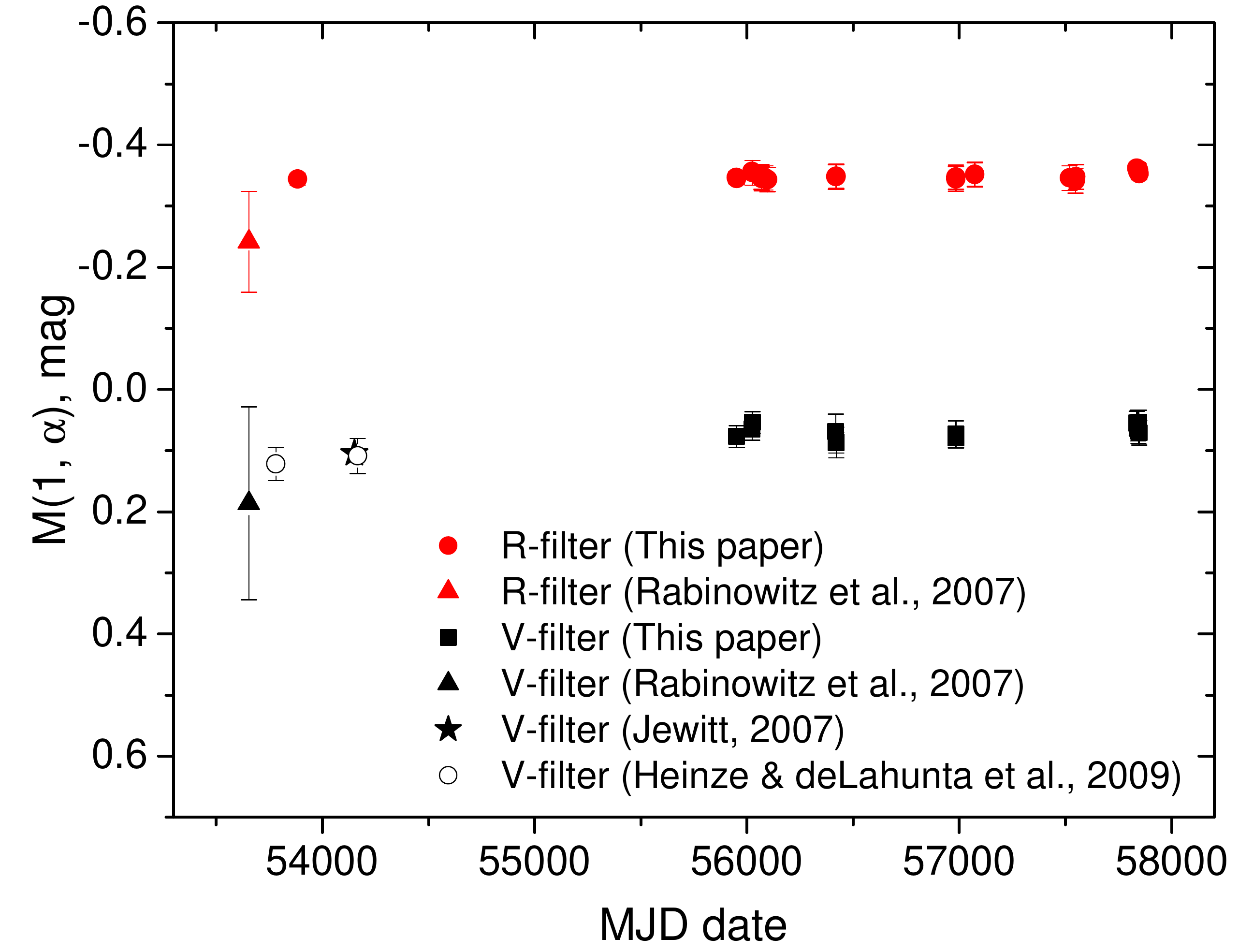}
        \caption{Magnitude vs modified Julian date (MJD) in V and R filters. The y-axis represents a mean reduced magnitude at the phase angle $\alpha$.}
        \label{mjd_mag}
\end{figure}

However, if rotational variability is indeed caused by an undiscovered satellite, its contribution in Makemake's absolute 
magnitude should be taken into account. Using the above-mentioned calculations of a possible satellite's size,
the absolute magnitude of Makemake should be fainter by at least 0.1~mag. 
This kind of correction was already performed in the case of Haumea \citep{Ortiz2017}. 
For Makemake it would imply that the geometric albedo should be about 10{\%} lower.
Also, we tested Makemake's brightness for long-term variability.
Namely, we were looking for changes in brightness lightcurve amplitude and absolute magnitude with time.
This information can help us make the assumption about the aspect of the observations and its evolution over ten years. 
The lightcurve amplitude of our data and the literature values remains very low.
The absolute magnitude of Makemake is also almost constant over the years (see Fig.~\ref{mjd_mag}).
Makemake's brightness seems to differ only in the \citet{Rabinowitz2007} data, 
whereas our data is constant within the errors and is in agreement with the \citet{Jewitt2007} results. 
It should be noted, however, that the photometry errors in \citet{Rabinowitz2007}
are rather large and for some data points well exceed the magnitude difference. 
In the graph we show the average magnitude value from \citet{Rabinowitz2007}, 
the error bars are the corresponding standard deviations of the data.
Because of the large distance from Earth, Makemake's aspect changes 
very slowly: since its discovery more than ten years ago, 
the ecliptic longitude has only changed  by about 11$^{\circ}$. 
This means that in order to notice some aspect changes from ground-based sites a much 
longer monitoring period is needed. At this point we can only exclude that
Makemake was reaching polar aspect during last ten years
because this would suggest a noticeable simultaneous decrease in
brightness amplitude (to its complete disappearance) and increase in absolute magnitude.

Overall, long and consistent monitoring is required in order to detect some aspect changes, which 
would lead to a better understanding of the true nature of Makemake's rotational period and to the physical and
orbital properties of its satellite(s).

\section*{Acknowledgements}

This research was partially based on data taken at the Sierra Nevada Observatory, 
which is operated by the Instituto de Astrofisica de Andalucia (CSIC). 
This research is also partially based on data taken at the German-Spanish Calar Alto observatory, 
which is jointly operated by the Max Planck Institute f{\"u}r Astronomie and the Instituto
de Astrofisica de Andalucia (CSIC).
J.L.Ortiz, R. Duffard, and P. Santos-Sanz acknowledges financial support from the 
State Agency for Research of the Spanish MCIU through the `Center of Excellence Severo Ochoa'
award for the Instituto de Astrofisica de Andalucia (SEV-2017-0709).
Funding from MINECO project AYA2017-89637-R is acknowledged.
Part of the research leading to these results has received funding from the European Union's Horizon 2020 Research and Innovation Programme, under Grant Agreement no 687378, as part of the project `Small Bodies Near and Far' (SBNAF).
R.Ya.I., V.R.A., and V.T.Z. are grateful to the Shota Rustaveli National Science Foundation
grant FR-18-1193 for  the partial financial support. 
D.P. has received funding from the European Union's Horizon 2020 research and innovation programme under the 
Marie Sklodowska-Curie grant agreement n. 664931.
I.V.R. and A.V.S. were partly supported by the programme BR05236322 of the
Ministry of Education and Science of the Republic of Kazakhstan and the 
scientific and technical program BR05336383 `Applied scientific research in the field of space activities'.
We thank D.~Chestnov and I.~Nikolenko for their help with obtaining observational data. 
We also want to thank S.~Lowry and A.~Heinze for providing feedback on the manuscript, which helped to improve the paper.

\bibliographystyle{aa} %
\bibliography{makemake_phot} %

\begin{appendix}
\section{Geometrical circumstances and magnitudes of Makemake}

\begin{table*}
        \centering
        \caption{Geometrical circumstances and magnitudes}
        \label{table_data}
        \begin{tabular}{p{2.3cm} p{1.0cm} p{1.0cm} p{1.0cm} p{1.0cm} p{1.0cm} p{1.4cm} p{1.5cm} p{0.9cm} p{0.9cm} c}
                \hline
                \hline
                UT date&r (AU)&$\Delta$ (AU)&$\alpha$ (deg)&$\lambda$ (deg)&$\beta$ (deg)&M (1, $\alpha$) (mag)&  Error$^{*}$ (mag)&        Filter  &$\Delta$T (h)&Obs.\\
                \hline
                2006 05 28.88&51.941&51.729&1.09&171.816&28.980&-0.364&0.01&  R &4.4&OSN\\
                
                2009 03 29.10&52.102&51.255&0.59&174.685&28.902&--&0.006&  R  & 6.4 &TNG\\
                
                2012 01 25.20&52.248&51.740&0.93&178.012&28.919&  0.956  & 0.02  &  B  &--&INT\\ 
                2012 01 25.20&52.248&51.740&0.93&178.012&28.919&  0.062  & 0.02  &  V  &--&INT\\
                2012 01 25.20&52.248&51.740&0.93&178.012&28.919&-0.357  & 0.01  &  R  &--&INT\\

                2012 01 26.00&52.250&51.727&0.92&178.001&28.922&  0.948  &  0.02   &  B  &--&INT\\
                2012 01 26.00&52.250&51.727&0.92&178.001&28.922&  0.077  &  0.02   &  V  &--&INT\\
                2012 01 26.00&52.250&51.727&0.92&178.001&28.922&-0.355&0.01&  R  &--&INT\\

                2012 04 07.98&52.260&51.442&0.64&177.491&29.001&  0.055  &  0.02  &   V  &6.5&CAO\\
                2012 04 08.00&52.260&51.442&0.64&177.491&29.001&-0.351&0.01&   R  &5.5&CAO\\

                2012 04 09.02&52.260&51.447&0.63&177.483&28.999&  0.054  &   0.02  &  V   &8.7&CAO\\ 
                2012 04 09.02&52.260&51.447&0.63&177.483&28.999&-0.354&0.02&  R  &8.2&CAO\\

                2012 05 21.90&52.261&51.874&1.03&176.696&28.98&-0.348&0.02&R    &4.3&Chuguev\\
                2012 05 22.84&52.266&51.886&1.03&176.690&28.972&-0.345 &0.02&R &3.1&Chuguev\\
                2012 06 08.83&52.268&52.128&1.10&176.630&28.823&-0.347&0.02&R   &1.6&Chuguev\\

                2012 06 18.84&52.269&52.277&1.11&177.913&28.737&-0.343&0.02&R   &3.1&Simeiz\\
                2012 06 19.84&52.269&52.292&1.11&177.915&28.737&-0.339&0.02&    R&2.5&Simeiz\\
                
                2013 05 06.86&52.312&51.720&0.90&178.791&28.678&0.044&0.03&  V  &--&OSN\\
                2013 05 06.87&52.312&51.720&0.90&178.791&28.678&-0.333&0.02&  R  &--&OSN\\

                2013 05 08.02&52.313&51.734&0.91&178.315&28.855&0.087&0.02&  V  &--&OSN\\
                2013 05 08.02&52.313&51.734&0.91&178.315&28.855&-0.359&0.02&  R  &--&OSN\\

                2013 05 09.10&52.313&51.734&0.92&178.311&28.851&0.012&0.03&  V  &--&OSN\\
                2013 05 09.10&52.313&51.734&0.92&178.311&28.851&-0.358&0.02&  R  &--&OSN\\

                2014 11 19.23&52.876&52.385&0.93&181.535&28.271&--&0.007&R&0.5&CAO\\

                2014 11 23.07&52.839&52.386&0.95&180.327&28.562&0.974&0.03&   B  &--&Terskol\\
                2014 11 23.07&52.839&52.386&0.95&180.327&28.562&0.023&0.02&   V  &--&Terskol\\
                2014 11 23.07&52.839&52.386&0.95&180.327&28.562&-0.359&0.02&R&0.8&Terskol\\
                2014 11 23.07&52.839&52.386&0.95&180.327&28.562&-0.623&0.03&   I  &--&Terskol\\

                2014 11 24.07&52.813&52.386&0.97&180.333&28.561&0.975&0.03&B&--&Terskol\\
                2014 11 24.07&52.813&52.386&0.97&180.333&28.561&0.049&0.02&V&--&Terskol\\
                2014 11 24.07&52.813&52.386&0.97&180.333&28.561&-0.358&0.02&R&2.1&Terskol\\
                2014 11 24.07&52.813&52.386&0.97&180.333&28.561&-0.602&0.03&I&--&Terskol\\

                2015 02 19.89&52.396&51.647&0.71&181.415&28.993&-0.352&0.02&R&7.2&Tian Shan\\
                2015 02 20.91&52.397&51.638&0.70&181.398&28.998&-0.361&0.02&R&5.7&Tian Shan\\
                2015 02 21.11&52.397&51.639&0.70&181.397&28.998&-0.360&0.02&R&3.2&OSN\\
                2015 02 23.98&52.395&51.616&0.67&181.348&29.011&-- &0.018&Clear&3.0&AbAO\\
                2015 02 25.04&52.395&51.609&0.66&181.331&29.015&-- &0.016&Clear&4.2&AbAO\\
                2015 02 25.95&52.396&51.603&0.65&181.314&29.019&-- &0.018&Clear&8.2&AbAO\\
                2015 02 26.95&52.397&51.598&0.65&181.297&29.022&-- &0.019&Clear&8.0&AbAO\\
                2015 02 27.95&52.397&51.592&0.64&181.280&29.026&-- &0.021&Clear&7.7&AbAO\\
                2015 03 12.13&52.399&51.541&0.55&181.060&29.056&--&0.009&R&4.9&OSN\\
                2015 03 13.13&52.399&51.538&0.55&181.041&29.058&--&0.06&R&4.5&OSN\\
                2015 03 15.17&52.399&51.534&0.54&181.004&29.060&--&0.06&R&1.8&OSN\\
                2015 03 16.10&52.399&51.534&0.54&180.085&29.061&--&0.07&R&2.8&CAO\\
                2015 03 27.88&52.399&51.531&0.54&180.757&29.060&--&0.022&Clear&7.0&AbAO\\
                2015 04 20.04&52.403&51.633&0.71&180.342&28.995&--&0.08&R&6.0&CAO\\
                
                2016 05 10.93&52.447&51.867&0.90&181.782&28.440&-0.339&0.02&R&1.5&CrAO\\
                2016 06 08.89&52.450&52.250&1.09&181.860&28.429&-0.346&0.02&R&3.0&CrAO\\
                2016 06 09.89&52.451&52.264&1.09&181.863&28.429&-0.348&0.02&R&3.4&CrAO\\
                
                2017 02 27.00&52.478&51.686&0.66&182.572&28.362&--&0.019&R&6.8&Simeiz\\
                2017 02 27.85&52.476&51.680&0.65&183.070&28.840&--&0.011&R&7.6&Tian Shan\\
                2017 02 28.04&52.478&51.680&0.65&182.575&28.362&--&0.008&Clear&4.5&Simeiz\\
                2017 02 28.99&52.478&51.674&0.64&182.578&28.361&--&0.009&Clear&7.2&Simeiz\\
                2017 03 22.93&52.480&51.605&0.53&182.637&28.355&--&0.007&Clear&8.1&Simeiz\\
                2017 03 23.92&52.480&51.605&0.53&182.640&28.355&--&0.008&Clear&8.3&Simeiz\\

                2017 03 24.03&52.480&51.605&0.53&182.640&28.355&0.056&0.02&V&--&OSN\\
                2017 03 24.03&52.480&51.605&0.53&182.640&28.355&-0.35&0.01&R&--&OSN\\

                2017 03 24.92&52.479&51.604&0.53&182.610&28.880&--&0.008&R&6.4&Simeiz\\
                2017 03 28.69&52.479&51.607&0.53&182.540&28.880&--&0.009&R&3.9&Tian Shan\\

                2017 03 29.06&52.481&51.609&0.53&182.653&28.354&0.054&0.02&V&--&OSN\\
        \hline
        \end{tabular}
\end{table*}
\begin{table*}
        \centering
        \begin{tabular}{p{2.3cm} p{1.0cm} p{1.0cm} p{1.0cm} p{1.0cm} p{1.0cm} p{1.4cm} p{1.5cm} p{0.9cm} p{0.9cm} c}
                Table continued&&&&&&&&&&\\
                \hline
                \hline
                2017 03 29.06&52.481&51.609&0.53&182.653&28.354&-0.356&0.01&R&--&OSN\\
                2017 03 30.05&52.481&51.610&0.54&182.656&28.353&0.074&0.02&V&--&OSN\\
                2017 03 30.05&52.481&51.610&0.54&182.656&28.353&-0.343&0.01&R&--&OSN\\
                2017 03 31.07&52.481&51.612&0.54&182.659&28.353&0.108&0.02&V&--&OSN\\
                2017 03 31.07&52.481&51.612&0.54&182.659&28.353&-0.345&0.01&R&--&OSN\\

                2017 04 02.07&52.481&51.616&0.53&182.664&28.353&0.094&0.02&V&--&OSN\\
                2017 04 02.07&52.481&51.616&0.53&182.664&28.353&-0.339&0.01&R&--&OSN\\

                2017 04 03.06&52.481&51.619&0.55&182.667&28.353&0.074&0.02&V&--&OSN\\
                2017 04 03.06&52.481&51.619&0.55&182.667&28.353&-0.359&0.01&R&--&OSN\\

                2017 04 03.99&52.481&51.621&0.56&182.669&28.352&0.051&0.02&V&--&OSN\\
                2017 04 03.99&52.481&51.621&0.56&182.669&28.352&-0.353&0.01&R&--&OSN\\

                2017 04 25.97&52.484&51.745&0.75&182.729&28.347&--&0.009&R&6.0&CAO\\
                2017 05 23.95&52.486&52.048&1.00&182.804&28.339&--&0.011&R&5.6&CAO\\
                2017 05 24.94&52.486&52.061&1.00&182.807&28.339&--&0.011&R&5.3&CAO\\
                2017 05 25.90&52.487&52.074&1.01&182.810&28.339&--&0.009&R&2.3&CAO\\
                2017 05 26.91&52.487&52.087&1.02&182.812&28.338&--&0.009&R&2.6&CAO\\
                \hline
        \end{tabular}
        \begin{flushleft}
        $^{*}$  The errors correspond to  absolute magnitude uncertainties when an absolute brightness was measured, 
        and to a differential photometry error otherwise.
        \end{flushleft}
\end{table*}

\end{appendix}

\end{document}